\documentclass[12pt]{article}

\usepackage{scicite}
\usepackage{graphicx}
\usepackage{siunitx}
\usepackage{amssymb}
\usepackage{times}
\usepackage{fdsymbol}
\usepackage{nameref}
\usepackage[version=4]{mhchem}
\usepackage{float}
\usepackage{bm}
\usepackage{upgreek}
\usepackage{textcomp}
\usepackage[dvipsnames]{xcolor}
\usepackage{comment}
\usepackage{setspace}

\usepackage{soul}

\title{Characterization of process-related interfacial dielectric loss in aluminum-on-silicon by resonator microwave measurements, materials analysis, and imaging}

\author
{Lert~Chayanun$^{1}$, Janka~Biznárová$^{1}$, Lunjie~Zeng$^{2}$, Per~Malmberg$^{3}$, 
\\
Andreas~Nylander$^{1}$, Amr~Osman$^{1}$, Marcus~Rommel$^{1}$, Pui~Lam~Tam$^{4}$,  
\\
Eva~Olsson$^{2}$, August~Yurgens$^{1}$, Jonas~Bylander$^{1}$, and Anita~Fadavi~Roudsari$^{1}$
\\
\normalsize{$^{1}$Chalmers University of Technology, Microtechnology and Nanoscience}\\
\normalsize{$^{2}$Chalmers University of Technology, Physics}\\
\normalsize{$^{3}$Chalmers University of Technology, Chemistry and Chemical Engineering}\\
\normalsize{$^{4}$Chalmers University of Technology, Industrial and Materials Science}\\
\normalsize{SE-41296, Gothenburg, Sweden}\\
\normalsize{$^\ast$Corresponding author's E-mail:  jankab@chalmers.se.}
}

\date{}

\begin{document} 


\baselineskip13pt

\maketitle 
\begin{abstract}
We systematically investigate the influence of the fabrication process on dielectric loss in aluminum-on-silicon superconducting coplanar waveguide resonators with internal quality factors ($Q_i$) of about one million at the single-photon level. These devices are essential components in superconducting quantum processors; they also serve as proxies for understanding the energy loss of superconducting qubits.
By systematically varying several fabrication steps, we identify the relative importance of reducing  loss at the substrate-metal and the substrate-air interfaces.
We find that it is essential to clean the silicon substrate in hydrogen fluoride (HF) prior to aluminum deposition. A post-fabrication removal of the oxides on the surface of the silicon substrate and the aluminum film by immersion in HF further improves the $Q_i$.
We observe a small, but noticeable, adverse effect on the loss by omitting either standard cleaning (SC1), pre-deposition heating of the substrate to \SI{300}{\celsius}, or in-situ post-deposition oxidation of the film’s top surface.
We find no improvement due to excessive pumping meant to reach a background pressure below \SI{6e-8}{\milli\bar}.
We correlate the measured loss with microscopic properties of the substrate-metal interface through characterization with X-ray photoelectron spectroscopy (XPS), time-of-flight secondary ion mass spectroscopy (ToF-SIMS), transmission electron microscopy (TEM), energy-dispersive X-ray spectroscopy (EDS), and atomic force microscopy (AFM).

\end{abstract}

\section*{\uppercase{Introduction}}

Superconducting coplanar waveguide (CPW) resonators~\cite{Zmuidzinas2012} in the microwave band are extensively used to read out the state of qubits in the circuit quantum electrodynamics (cQED) architecture~\cite{Wallraff2004,Frunzio2005,Wang2009,Sage2011, Megrant2012}. They can also be used as tools to investigate the loss in superconducting qubits~\cite{Dunsworth2017,Burnett2018,Sage2011}. Several loss mechanisms can adversely affect the performance of quantum circuits, including charged two-level systems (TLS), spins, non-equilibrium quasiparticles, radiation, and magnetic vortices \cite{McRae2020,Alto2022,Crowley2023}. Among these, TLS dominate the loss at millikelvin temperatures and single-photon-power excitations, i.e.\@ at the operating conditions of a  superconducting quantum processor~\cite{Martinis2005}. 
TLS loss originates in dielectric amorphous materials~\cite{Müller2019}, mainly within the oxide layers located at the substrate-air (SA), metal-air (MA), and substrate-metal (SM) interfaces of the device. Coupling to a bath of TLS, or to spurious, individual TLS, results in qubit decoherence and resonator loss~\cite{Martinis2005,Gao2008}.\par
The TLS loss can be mitigated by refining the fabrication process. Different wet and dry etching processes have been employed to clean the silicon (Si) substrate before the metal deposition to decrease the loss at the SM interface~\cite{Wisbey2010,Bruno2015,Earnest2018,Fritz2019}. The choice of silicon (and sapphire) as popular substrates is due to their low loss tangent values~\cite{OConnell2008,Wang2015}, which is essential in minimizing the loss in areas with high field concentrations. Various etching methods, including trenching of the substrate, have been explored to transfer the resonator pattern onto the metal layer while maintaining a low-loss SA interface~\cite{Barends2010,Sandberg2012,Vissers2012,Richardson2016,Calusine2018}. Both the SA and MA interfaces can be cleaned post processing, before loading the sample into the cryostat for measurement~\cite{Lock2019,Kowsari2021}. In addition, the metal surface can be oxidized in-situ after the deposition by filling the chamber with pure oxygen~\cite{Burnett2018}. Passivation of the superconducting metal at the MA interface has also been explored \cite{Zheng2022}. We have combined some of these processes in a standard fabrication process for aluminum-on-silicon (Si-Al), discussed below, and it yields CPW resonators and qubits with TLS-limited internal quality factor $Q_\text{i} \approx 1\times 10^6$ or relaxation time $T_1\approx \SI{100}{\micro\second}$~\cite{Burnett2019,Osman2021,Kosen2022}. The choice of aluminum as the superconducting material originates from its availability and the vast knowledge about its chemistry and fabrication techniques. A recent demonstration of Al transmon qubits with average relaxation time  $T_1 > \SI{250}{\micro\second}$ highlights the enduring potential of aluminum as a platform for superconducting circuits~\cite{Biznarova2023}. \par
This report presents a systematic evaluation of our CPW resonator fabrication process. We fabricate control resonators based on the standard recipe previously developed in our group~\cite{Burnett2018}, and also other resonators using recipes that deviate from the standard by only one step. We determine the resonators' $Q_\text{i}$ and quantify the associated TLS losses in each fabrication step. In parallel, we use atomic force microscopy (AFM) to study the film surface, and transmission electron microscopy (TEM) to investigate the material structures at different interfaces of the resonators. 
Furthermore, we use a complementary set of techniques for compositional depth profiling, including X-ray photoelectron spectroscopy (XPS), energy dispersive X-ray spectroscopy (EDS) and time-of-flight secondary ion mass spectroscopy (ToF-SIMS).
XPS is suitable for exploring the chemical composition and the chemical state of the selected elements, while the ability of ToF-SIMS to detect traces of individual elements and molecules with part-per-million accuracy can unfold the elemental depth profile of the device. We find that the most critical step to minimize the loss at the substrate-metal interface is the initial removal of the native oxide off the surface of the silicon wafer in hydrofluoric acid (HF). The substrate-air interface loss can also be reduced temporarily, post fabrication, by the same acid treatment.

\section*{\uppercase{Method}}
\begin{figure}
    \centering
    \includegraphics[width=8.5cm]{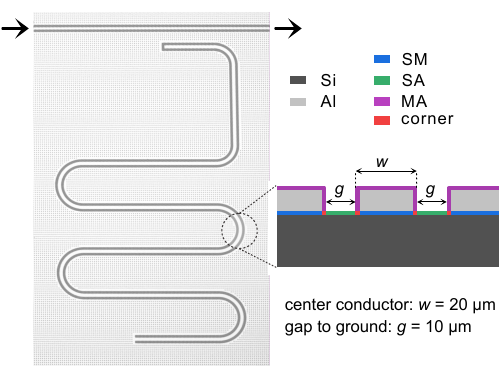}
    \caption{\small{\label{fig:Device} The device and its materials interfaces. The micrograph shows the quarter-wavelength CPW resonator, coupled capacitively to an input-output transmission line and surrounded by an array of $2\times\SI{2}{\micro \meter\squared}$ flux-trapping holes. The cartoon illustrates a CPW cross-section, highlighting the interfaces that contribute to the dielectric loss (not to scale). }
    }
\end{figure}

%
The internal quality factor ($Q_\text{i}$) is used as a figure of merit to quantify the performance of a CPW resonator. We obtain $Q_\text{i}$ of the resonator from the microwave forward transmission ($S21$) measurement of a quarter-wavelength resonator in a notch configuration relative to the feedline, see Fig.~\ref{fig:Device}, using a routine circle fit~\cite{Probst2015}. The inverse of $Q_\text{i}$ is referred to as the loss tangent,
\begin{equation}\label{eq:1_Q-losstangents}
    1/Q_\text{i} = \text{tan}\delta \approx \delta \,,
\end{equation}

for $\text{tan}\delta \ll \text{1}$ \cite{McRae2020}. $\delta$ is a combination of different losses, $\delta = \delta_\text{TLS} + \delta_\text{other}$, where $\delta_\text{TLS}$ is the loss due to TLS and $\delta_\text{other}$ is the sum of other loss mechanisms. According to the TLS model \cite{Gao2008}, $\delta_\text{TLS}$ is written as a function of power and temperature
\begin{equation}\label{eq:delta_TLS}
    \delta_\text{TLS} = F\delta_\text{TLS}^0 \frac{\text{tanh}\left(\hbar \omega_0/2k_\text{B}T\right)}{\left(1+ \langle n \rangle / n_\text{c} \right)^\beta} \,,
\end{equation}

where $F$ is the filling factor, defined as the ratio of the electric field stored in the TLS material to the total stored electric field, $\delta_\text{TLS}^0$ is the intrinsic TLS loss, and $\omega_0$ is the resonator frequency. $T$ represents the temperature, $\langle n \rangle$ is the average number of photons stored/circulating in the resonator, $n_\text{c}$ is the critical photon number to saturate a single TLS, and  $\beta \le 0.5$ is the TLS saturation rate with photon power~\cite{Burnett2017}.  $\hbar$ and  $k_B$ are the reduced Planck and Boltzmann constants, respectively. Combining Eq.~\ref{eq:1_Q-losstangents} and Eq.~\ref{eq:delta_TLS} and assuming $\text{tanh}({\hbar \omega_0}/{2k_\text{B}T})\approx 1$, we get 
\begin{equation}\label{eq:1_Q-final}
    1/Q_\text{i} = F\delta_\text{TLS}^0 \frac{1}{\left(1+\langle n \rangle / n_\text{c} \right)^\beta} + \delta_\text{other} \,,
\end{equation}

which we use to quantify the TLS loss $F\delta_\text{TLS}^0$ of the superconducting resonators.\par
%
For our experiments, we used high-resistivity intrinsic Si (100) wafers and aluminum superconducting films. The standard fabrication process started with wafer cleaning, by submerging the wafer in a mixture of ammonium hydroxide, hydrogen peroxide, and deionized (DI) water (\ce{NH4OH}:\ce{H2O2}:\ce{H2O}, 1:1:5) -- known as Standard Clean\,1 (SC1) -- at \SI{80}{\celsius} for \SI{10}{\minute}. 
The wafer was then rinsed with DI water and subsequently dipped in a
2\% aqueous solution of hydrofluoric acid for \SI{1}{\minute},
followed by another DI water rinse. Within a few minutes we loaded the cleaned wafer into the load-lock chamber of the evaporator \textit{Plassys\,MEB550S} to minimize the re-oxidation of the Si surface \cite{Morita1990}. Next, we heated the substrate, \textit{in-situ}, to \SI{300}{\celsius} for \SI{10}{\minute}. After about \SI{20}{\hour}, the wafer cooled down to room temperature ($\sim$\SI{20}{\celsius}), and the chamber base pressure reached \SI{4e-8}{\milli\bar}. We evaporated \SI{150}{\nano\meter} of Al at the deposition rate of \SI{1}{\nano\meter\per\second}, followed by static oxidation of the Al surface at the pressure of \SI{10}{\milli\bar} for \SI{10}{\minute}.\par
Following the film deposition, the resonators were fabricated by means of optical lithography and wet etching and measured in a dilution refrigerator at about \SI{10}{\milli\kelvin}. Besides the standard sample (\#1), we measured five additional samples (\#2-6), where for systematic evaluation, we fabricated each sample by varying only one step of the standard process. The list of the samples and the type of variation are summarized in Table~\ref{tab:Samples1-6-info}. We fabricated sample \#1 using the standard process, while for sample \#2 we did not remove the native oxide of the silicon substrate in HF prior to metal evaporation. For samples \#3, \#4, and \#5, we skipped the SC1 step, the substrate preheat, and the in-situ oxidation of aluminum, respectively, and finally, we prepared sample \#6 using a short (\SI{4}{\hour}) pumping time (instead of overnight) prior to metal evaporation. \par

\begin{table}[b]
\caption{\small{\label{tab:Samples1-6-info} Summary of samples \#1-6 (evaporated Al films). The table includes the average $Q_\text{i}$ at $\langle n \rangle = 1$, $\beta$, $F\delta_\text{TLS}^0$ and $\delta_\text{other}$ as fit parameters to the TLS model, Eq.~\ref{eq:1_Q-final}.}}

\small
\footnotesize
\renewcommand{\arraystretch}{1.1}
\centering\begin{tabular}{l c c c c c c}
\\[-1.8ex]
\hline \hline
\\[-1.8ex]
     & \textbf{Sample 1} & \textbf{Sample 2} & \textbf{Sample 3} & \textbf{Sample 4} & \textbf{Sample 5} & \textbf{Sample 6} \\
     & \textbf{(\#1)} & \textbf{(\#2)} & \textbf{( \#3)} & \textbf{(\#4)} & \textbf{(\#5)} & \textbf{(\#6)} \\[3pt]
   Process variation & Standard & No HF & No SC1 & No preheat & No in-situ oxidation & Short pumping \\[3pt]
   \hline \\[-1.8ex]
   $Q_\text{i}$\,($\langle n \rangle = 1$) & $1.1\times10^6$ & $0.24\times10^6$ & $0.95\times10^6$ & $0.8\times10^6$ & $0.92\times10^6$ & $1.1\times10^6$ \\[3pt]
			
   $\beta$ & $0.22$ & $0.29$ & $0.27$ & $0.20$ & $0.27$ & $0.24$ \\[3pt]
									
   $F\delta_\text{TLS}^0$ & $0.87\times10^{-6}$ & $4.4\times10^{-6}$ & $1.0\times10^{-6}$ & $1.4\times10^{-6}$ & $0.97\times10^{-6}$ & $0.87\times10^{-6}$ \\[3pt]

   $\delta_{other}$ & $2.3\times10^{-7}$ & $2.3\times10^{-7}$ & $2.0\times10^{-7}$ & $2.1\times10^{-7}$ & $2.1\times10^{-7}$ & $2.4\times10^{-7}$ \\[2pt]
\hline \hline
   
\end{tabular}

\end{table}

Subsequently, we investigated the influence of varying the conditions of the film deposition itself. For wafer \#7 (includes samples \#7a-c), we followed the same conditions as for sample \#1, with the exception that the SC1 cleaning was omitted and the Al film was sputtered instead of evaporated: after cleaning the substrate using HF dip, the wafer was immediately transferred to the heated load-lock (\SI{80}{\celsius}) of a sputter tool (\textit{DCA\,MTD620}), which was pumped down for \SI{40}{\minute} until a pressure below \SI{5e-7}{\milli\bar} was reached. Thereafter, the wafer was transferred to the deposition chamber. Here the substrate was heated to \SI{300}{\celsius} for \SI{10}{\min} and pumped for \SI{16}{\hour}, at which point the base pressure of the deposition chamber was reached at \SI{2.2e-8}{\milli\bar}. Al was then deposited by direct-current (DC) magnetron sputtering in argon (Ar) plasma, followed by in-situ oxidation of the film surface. In the Supplementary materials, we also present an investigation into the effects on the resonator $Q_i$ resulting from the deposition rate, substrate temperature during deposition, and high-temperature substrate preheating.\par
We used three samples, \#7a--c, to investigate the effect of a short dip in 2\% HF post fabrication, with the aim to remove the surface oxides of both the Al film and the Si substrate, and with them any adsorbates left behind by the fabrication process. This dip was only \SI{15}{s} long due to the tendency of Al to be etched in this solution, and the samples were rinsed in DI water afterwards. The effect of this procedure is shown on sample \#7b, while \#7a is kept as the control. Sample \#7c underwent a vapor HMDS deposition at \SI{100}{\celsius} immediately after HF treatment, where the hydrophobic HMDS monolayer deposited this way could be expected to protect the Si from re-oxidizing~\cite{Bruno2015}. More details on samples \#7a--c are found in Table~\ref{tab:Sample-info-DCA}. Due to the tendency of both Al and Si to regrow oxides in ambient conditions, samples \#7b and \#7c were mounted inside the dilution refrigerator, with the pump-down preceding the cool-down started within 2.5~h after the HF dip.\par

The XPS experiment was performed using a \textit{PHI\,5000\, VersaProbe\,III} system equipped with a monochromatic aluminum X-ray source (i.e. Al\,K${\alpha}$). To aid in depth profiling, we performed a step-wise iterative procedure of Ar$^+$ ion sputtering to etch away the Al layer followed by an XPS measurement, until the SM interface was reached. We scanned selected binding-energy ranges based on the elements of interest, including oxygen (O\,1s), aluminum (Al\,2p), and silicon (Si\,2p),  in order to conduct a qualitative analysis of the chemical state(s) of individual elements. 

TEM bright field (BF) imaging was performed using a \textit{FEI Tecnai T20} microscope operated at 200~kV.  TEM cross-section samples were prepared by a \textit{FEI Versa 3D} focus ion beam -  scanning electron microscope (FIB-SEM). Scanning transmission electron microscopy (STEM) energy dispersive X-ray spectroscopy (EDS) analysis was carried out using a \textit{JEOL monochromated ARM 200F} TEM, which is equipped with a Schottky field emission electron source, a double-Wien monochromator, a probe Cs corrector, an image Cs corrector, as well as a double silicon drift detector (SDD) for EDS. 

Additionally, we performed the ToF-SIMS measurement using an \textit{IONTOF\,5} system, with a bismuth ion (Bi$^+$) gun for imaging and a cesium ion (Cs$^+$) sputter gun for depth profiling. The mass spectrum of the secondary ions emitted from the sample was collected at each sputtering step and converted to an elemental depth profile. Note that the XPS and ToF-SIMS measurements were performed either on the chips from the same wafers as the resonators were fabricated on, or on chips from wafers that were prepared following identical cleaning and film growth procedures.  \par
%
\begin{table}[]
\caption{\small{\label{tab:Sample-info-DCA} List of samples \#7a-c (sputtered Al films). Samples \#7a is a control. Samples \#7b and \#7c were treated with HF post fabrication, with the addition of an HMDS step in \#7c. The table includes the average $Q_\text{i}$ at $\langle n \rangle \sim 10$ , $\beta$, $F\delta_\text{TLS}^0$, and $\delta_\text{other}$ for the TLS model fit for every sample, as shown in Fig.~\ref{fig:resonators_sputtered_andHF}(a). }}

\small
\footnotesize
\renewcommand{\arraystretch}{1.1}
\centering\begin{tabular}{l c c c}
\\[-1.8ex]
\hline \hline
\\[-1.8ex]
     & \textbf{Sample 7a} & \textbf{Sample 7b} & \textbf{Sample 7c} \\
     & \textbf{(\#7a)} & \textbf{(\#7b)} & \textbf{(\#7c)} \\[3pt]
   Process variation & Sputter reference & HF & HF + HMDS \\[3pt]
   \hline \\[-1.8ex]
   $Q_\text{i}$\,($\langle n \rangle \sim 10$) & $1.1\times10^6$ & $2.2\times10^6$ & $2.1\times10^6$ \\[3pt]
			
   $\beta$ & $0.36$ & $0.33$ & $0.36$ \\[3pt]
					
   $F\delta_\text{TLS}^0$ & $0.86\times10^{-6}$ & $0.27\times10^{-6}$ & $0.39\times10^{-6}$ \\[3pt]

   $\delta_{other}$ & $1.5\times10^{-7}$ & $2.0\times10^{-7}$ & $1.7\times10^{-7}$ \\[2pt]
\hline \hline
   
\end{tabular}

\end{table}

\section*{\uppercase{Results and Discussion}}

\begin{figure}[t]
    \centering
    \includegraphics[width=8.5cm]{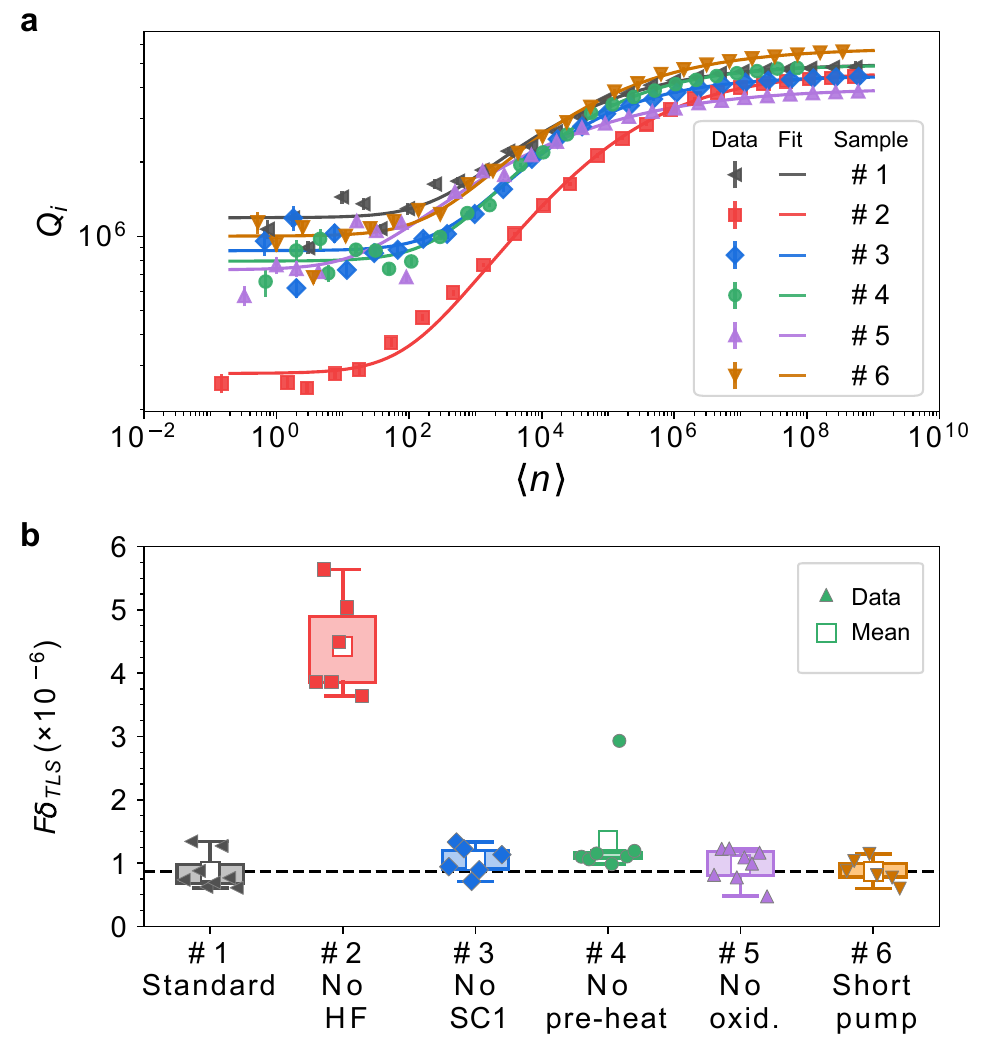}
    \caption{\small{\label{fig:resonators1-6} Resonator loss as fitted to the TLS model (Eq.~\ref{eq:1_Q-final}) for samples \#1-6, with process variations listed in Table~\ref{tab:Samples1-6-info}. \textbf{a)} $Q_\text{i}$ vs.\@ circulating power. The resonance frequency is $\sim$~\SI{4.4}{\giga\hertz}. \textbf{b)} Box plot of TLS loss. Each data point represents the loss of an individual resonator.
    }}
\end{figure}

Each sample contains up to eight resonators, whose frequencies range between 4 and 8~GHz. We measured the power-dependent $S_{21}$ using a vector network analyzer (VNA) and calculated $F\delta_\text{TLS}^0$ by fitting the results with the TLS loss model described in Eq.~\ref{eq:1_Q-final}. Figure~\ref{fig:resonators1-6}(a) presents the average $Q_\text{i}$ as a function of $\langle n \rangle$ for samples \#1-6. For samples fabricated with the standard process, on average, we find $Q_\text{i} = 1.1 \times 10^6$ when $\langle n \rangle = 1$, and $F\delta_\text{TLS}^0 = 0.87 \times 10^{-6}$. For the other samples the value of $Q_\text{i}$ together with the fitting parameters of the TLS loss model (Eq.~\ref{eq:1_Q-final}) are found in Table~\ref{tab:Samples1-6-info}. The parameter $F\delta_\text{TLS}^0$ from the fit is presented as a box plot in Fig.~\ref{fig:resonators1-6}(b), where each point in this plot is from an individual resonator. The loss due to other parameters $\delta_\text{other}$, also presented in Table~\ref{tab:Samples1-6-info}, is about $0.2 \times 10^{-6}$ for all samples. \par
\begin{figure}[t!]
    \centering
    \includegraphics[width=16.5cm]{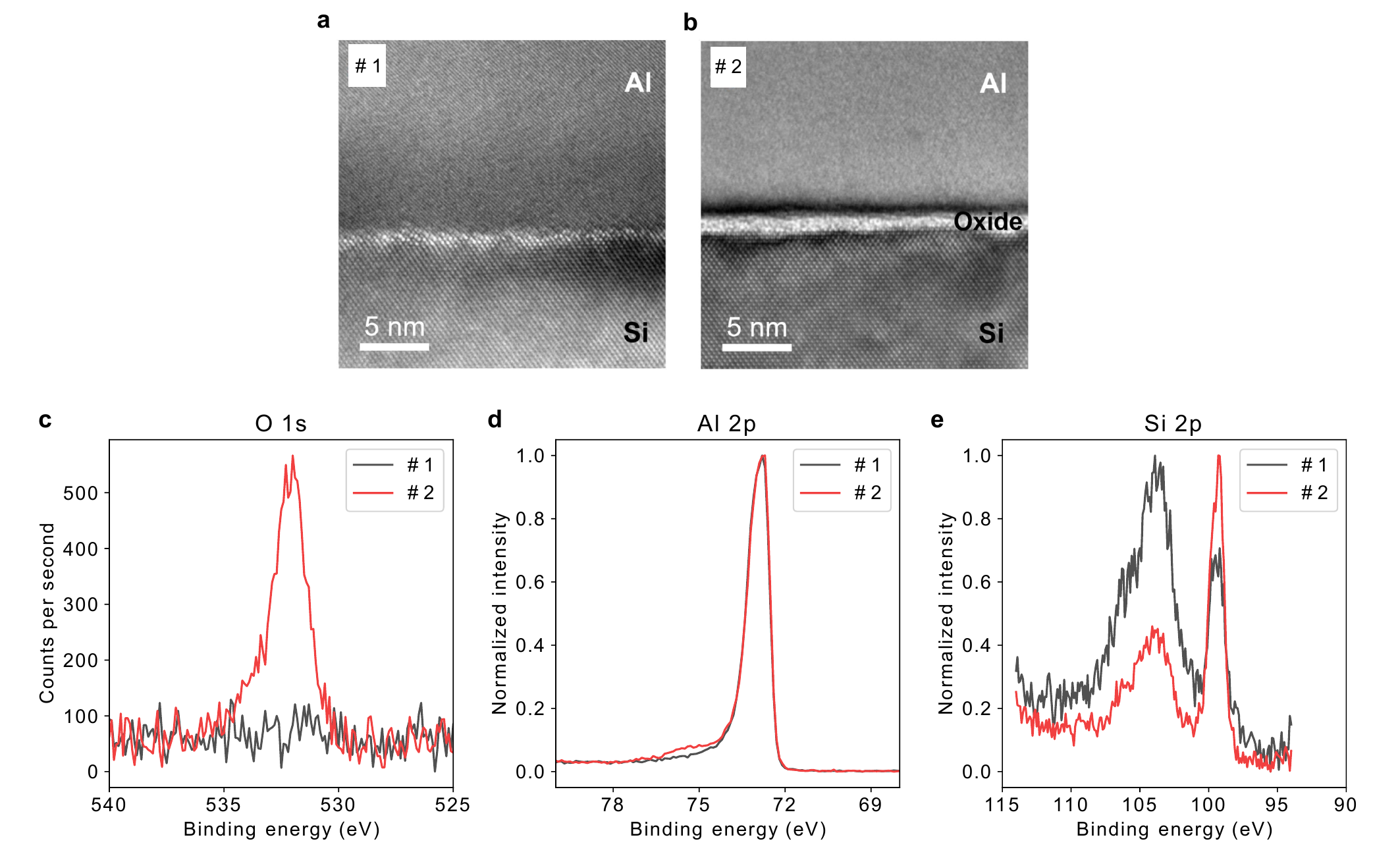}
    \caption{\small{\label{fig:TEM-XPS} TEM images and XPS spectra comparing the SM interface of the standard sample (\#1) and the sample without the HF clean (\#2). \textbf{a)}  The TEM BF image of sample \#1 shows a clean, oxide-free interface. Al and Si lattices directly contact each other at the interface at the atomic scale, indicating a clean interface without an amorphous interface layer. The bright contrast at the interface is due to Fresnel fringe contrast. \textbf{b)} The TEM of sample \#2 shows an oxide layer at the SM interface with a thickness of about \SI{1.5}{\nano\meter}.}  \textbf{c)} -- \textbf{e)} XPS spectra of O\,1s, Al\,2p, and Si\,2p at the SM interface resulting from the standard-process (black) and no-HF (red) samples.
    }
\end{figure}

The highest level of TLS loss is observed in sample \#2 (no HF), a 5-fold increase compared to that of the standard sample (\#1), see Fig.~\ref{fig:resonators1-6}(b). 
Using TEM, we find a clean SM interface of the standard sample, whereas there exists a layer of $\sim$\SI{1.5}{\nano\meter}-thick oxide at the SM interface of \#2, Figs.~\ref{fig:TEM-XPS}(a--b). 
The XPS spectra for O\,1s, Al\,2p, and Si\,2p at the SM interface of samples \#1 and \#2 are shown in Figs.~\ref{fig:TEM-XPS}(c--e), respectively. Although the oxide layer at the SM interface of \#2 was supposedly comprised of native oxide atop the Si substrate ($\text{SiO}_2$), the oxidation-reduction reaction at room temperature between Al and $\text{SiO}_2$ has turned the oxide into an aluminum oxide type \cite{Bauer1980}. The signature of this oxide is observed through the XPS spectra of O\,1s and Al\,2p. The Al\,2p spectrum shows that metallic Al, with the binding energy position at \SI{72.8}{\eV}, is the dominating chemical state, while a small shoulder positioned at \SI{75.3}{\eV} corresponds to the oxidized Al\,(III) state. The major peak of O\,1s at \SI{532}{\eV} gives an energy difference of \SI{456.8}{\eV} with Al\,(III). This implies that the interface oxide is mostly in the form of aluminum oxide. Still, as the oxygen content is rather low and close to the XPS detection limit, i.e. 1.0~at.\%, it is difficult to determine exactly in which form the oxide has developed, see Figs.~\ref{fig:TEM-XPS}(c--d). The characteristic peak of Si\,2p at \SI{99.4}{\eV} indicates that silicon is mostly in its elemental state. The secondary peak at \SI{104.0}{\eV} refers mostly to the Al\,2p plasmon loss peak instead of the overlapping Si\,(IV) state, with reference to both the energy difference and the area ratio compared to the metallic Al\,2p peak, see Figs.~\ref{fig:TEM-XPS}(d--e). \par
%

\begin{figure}[t]
    \centering
    \includegraphics[width=9.5cm]{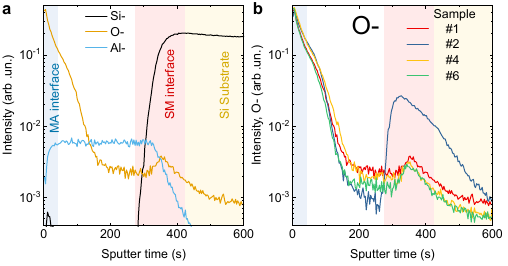}
    \caption{\small{\label{fig:ToF-SIMS} ToF-SIMS. \textbf{a)} Depth profile of the standard-process sample (\#1) characterized by ToF-SIMS. The depth is proportional to the time during which the sample was etched by the Cs$^+$ sputter beam. The MA and SM interfaces are highlighted with blue and red, respectively, while the Si substrate is in yellow. \textbf{b)} The depth profile of O from sample \#1 (standard process), \#2 (no HF), \#4 (no preheat), and \#6 (short pumping). 
    }}
\end{figure}

The XPS spectra indicate no trace of O\,1s in sample \#1 at the Si-Al interface, see Fig.~\ref{fig:TEM-XPS}(c). However, the depth profile of \#1 from the ToF-SIMS measurement in Fig.~\ref{fig:ToF-SIMS}(a) features a small peak from oxygen at the SM interface. Since the oxide was not observed in TEM either, we speculate that the trace oxide in \#1 is either local or in the form of a discontinuous layer along the grain boundaries~\cite{Biznarova2023}. The oxygen signal at the SM interface in \#2 is more intense and agrees well with the TEM and XPS results, see Fig.~\ref{fig:ToF-SIMS}(b). The intensity of the O peak in \#2 is almost an order of magnitude higher than that of the other samples, in agreement with the relatively higher concentration of O at the SM interface of this sample and its correspondingly high TLS loss level, see Fig.~\ref{fig:resonators1-6}.\par
Surface treatment with SC1 and HF has been a prominent cleaning method in the semiconductor industry for several decades~\cite{Gale2018}. The SC1 solution cleans the surface of the Si substrate from most organics and some metallic contamination by trapping them into an oxide layer on the surface, which is then removed by HF. If SC1 is not performed, some contamination remains  on the substrate surface, resulting in 15\% 
higher $F\delta_\text{TLS}^0$ in \#3 compared with \#1 (Fig.~\ref{fig:resonators1-6}(b)), even though the oxide layer, as the main source of TLS loss, has already been removed by the HF solution.\par
Another process extensively used to ensure a good quality SM interface is the heating of the substrate under vacuum~\cite{Fritz2018,Richardson2020,Cerofolini1997}. The elevated temperature of the substrate can desorb moisture and volatile molecules from the surface~\cite{Naik2018}. In some cases, the substrate undergoes annealing temperatures of above \SI{700}{\celsius} for surface reconstruction to achieve a better atomic transition between the metal and substrate~\cite{Fritz2019,McSkimming2017}. In our experiments, the sample with no preheating process (\#4) showed a 60\% 
increase in $F\delta_\text{TLS}^0$. \par
\begin{figure}[t]
    \centering
    \includegraphics[width=9.5cm]{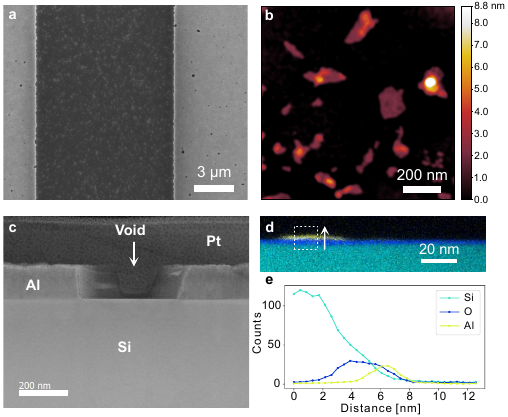}
    \caption{\small{\label{fig:SEM-AFM} SEM, AFM, TEM and EDS of sample \#5 (no in-situ oxidation of Al). \textbf{a)} In the SEM, voids appear as dark spots in the Al film (silver). Besides, flakes are observed in the etched Al area on Si (gray). \textbf{b)} AFM image of the flakes with the lateral size ranging from a few tens of nanometers to \SI{200}{\nano\meter} and thickness of about \SI{2}{\nano\meter}. \textbf{c)} TEM BF image showing a void in the Al film. The Pt film was deposited on the sample during TEM cross-section sample preparation to protect the structure. \textbf{d)} Composite EDS composition map of Al (yellow), O (blue), and Si (cyan) in an area where the Al layer has been etched. The map was obtained using Al-K, O-K, and Si-K EDS signals. The dashed-square line and arrow indicate the area and direction where the EDS intensity profiles were extracted. The profiles are shown in \textbf{e)}.}
    }
\end{figure}

The TLS loss of sample \#5 is just marginally higher (10\%) 
than the standard sample, see Fig.~\ref{fig:resonators1-6}(b), and one may conclude that in-situ oxidation of the Al film is unnecessary. However, upon SEM inspection of \#5, we found dark spots distributed over the Al film -- resembling voids -- as shown in Fig.~\ref{fig:SEM-AFM}(a). The TEM image of Fig.~\ref{fig:SEM-AFM}(c) shows the cross section of one of these voids.  
For further investigation, we prepared a second Al film on the Si substrate where we skipped the in-situ oxidation after evaporating the film. We observed no voids in the Al immediately after deposition. Next we heated a sample to \SI{160}{\celsius} for \SI{5}{\minute} to create a similar condition to that of baking the photoresist during resonator fabrication. We inspected the sample daily; the voids appeared on the piece after about two days.\par
%
The SEM image of sample \#5 shows residues on the silicon substrate within the CPW gap where Al was wet etched, see Fig.~\ref{fig:SEM-AFM}(a). These residues remain on the silicon surface after wet etching of the Al layer using aluminum etchant (the mixture of phosphoric, nitric,
and acetic acids), even if we etch a sample that is not exposed to heating after metal deposition. A longer etch time does not remove the residues either. The lateral size of these residues varies from a few tens of nanometers up to \SI{200}{\nano\meter}, and the thickness is measured to be approximately \SI{2}{\nano\meter}, as shown in the AFM image of Fig.~\ref{fig:SEM-AFM}(b). Cross-sectional EDS analysis in TEM (Figs.~\ref{fig:SEM-AFM}(d--e) and Supplementary Materials) show the composition distribution of a residue flake in the etched Al area. There is an intermixing of Al, Si and O in the flake. \par
Due to the thermal expansion mismatch between Al and Si, heating or cooling can generate stress in the Al film ($\alpha_\text{Al}$ = 23.1 $\times10^{-6}$$\,\,^{\circ}\text{C}$${^{-1}}$ and $\alpha_\text{Si}$ = 2.6 $\times10^{-6}$$\,\,^{\circ}\text{C}$${^{-1}}$). The textured (polycrystalline) form of evaporated Al and its low melting point make it prone to recrystallization. Stress can promote the recrystallization and formation of defects in the film. The development of hillocks and voids in pure Al films on Si (or even silicon oxide) has been a known challenge in the CMOS industry \cite{Plummer_1999_book}. The oxide on the surface of Al might help to keep the grains together and partially prevent the creation of defects. We speculate that different types of aluminum oxide grow atop the Al film as a result of in-situ oxidation (exposure to pure oxygen at low pressure), or ambient air oxidation (exposure to oxygen and moisture at atmospheric pressure). Oxidation in the ambient air might introduce more defects and porosity in the aluminum oxide, making it less effective in protecting the film. This can result in the formation of the voids and defects in the Al film. \par
%

With the shorter pumping time used for sample \#6, the base pressure and the substrate temperature prior to deposition were \SI{6e-8}{\milli\bar} and \SI{50}{\celsius} (passively cooled down from \SI{300}{\celsius}), respectively. The $Q_\text{i}$, and TLS loss of \#6 are comparable to those of \#1 (Fig.~\ref{fig:resonators1-6}). The ToF-SIMS results of \#6 show only a negligible change of O intensity  at the SM interface in Fig.~\ref{fig:ToF-SIMS}(b), meaning that longer pumping time is not necessary. \par

\begin{figure}[t]
    \centering
    \includegraphics[width=8.5cm]{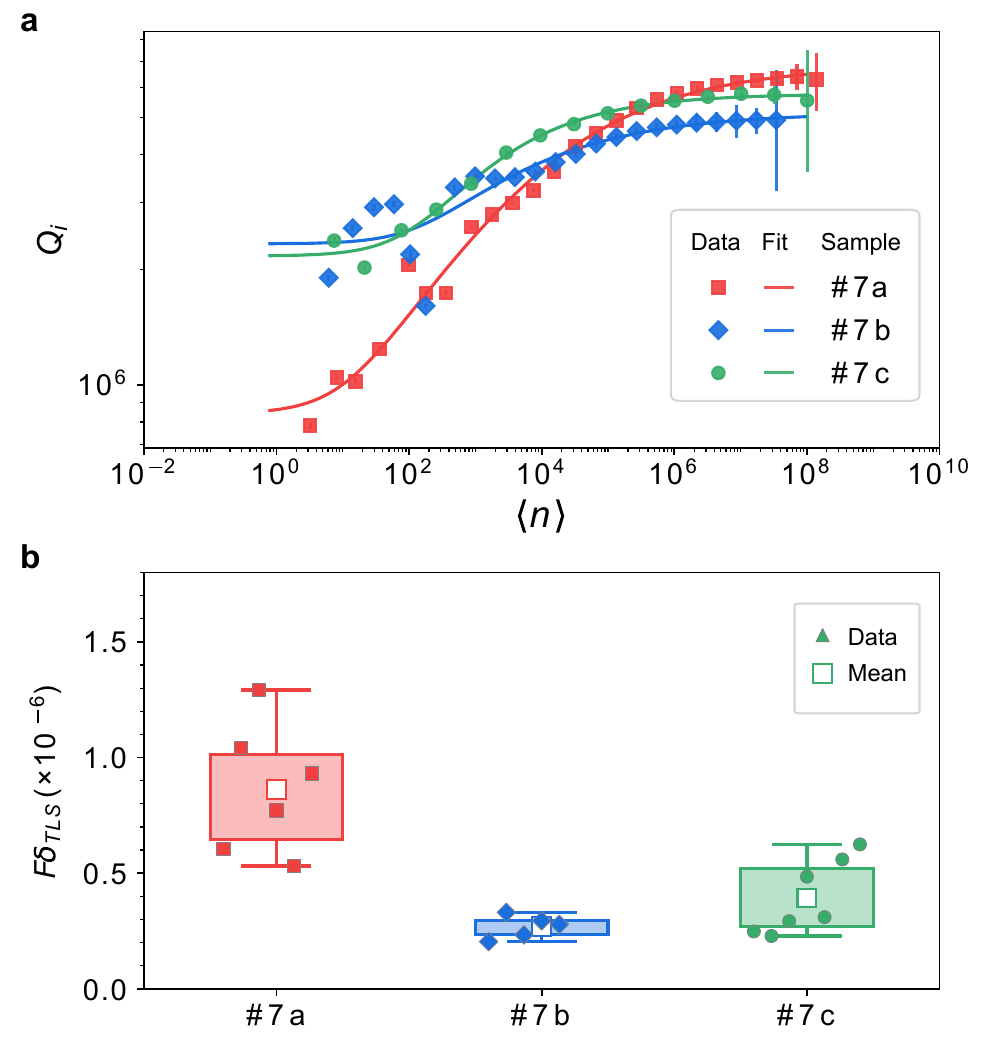}
    \caption{\small{\label{fig:resonators_sputtered_andHF} Internal quality factor ($Q_\text{i}$) and losses determined from the TLS model fits for the samples with a sputtered Al film. \textbf{a)} $Q_\text{i}$ vs.\@ average number of photons for the resonators fabricated with the process variations listed in Table~\ref{tab:Sample-info-DCA}. The resonance frequency is $\sim$~\SI{4.8}{\giga\hertz}. \textbf{b)} Box plot of TLS loss from the fits to the TLS loss model for all of the resonators on each sample. Each data point represents the loss of an individual resonator.
    }}
\end{figure}

For the resonators etched out of a sputtered Al film, we achieved a comparable quality factor to those on the evaporated film, see Fig.~\ref{fig:resonators_sputtered_andHF} and Table~\ref{tab:Sample-info-DCA}. The average TLS loss $F\delta_\text{TLS}^0$ obtained for sample \#7a is about $0.86\times10^{-6}$, which is comparable to that of \#1. This is despite the increased surface roughness of the sputtered film compared to the evaporated film, which may increase the loss at the MA interface. The other, non-TLS related loss mechanisms are lower for the sputtered film \#7a (smaller $\delta_{other}$). Since $\delta_{other}$ is a combination of several loss mechanisms, pinpointing the exact cause of the lower loss needs further study. A fraction of this may be an effect of lowered quasiparticle loss due to an increased superconducting gap in the sputtered film, as the superconducting critical temperature $T_c$, collected from 4-point probe DC transport measurements, is \SI{1.2}{\kelvin} for the sputtered film, and \SI{0.91}{\kelvin} for the evaporated (standard) film. In contrast, the residual-resistance ratio RRR is 3 for the sputtered film and 8 for the evaporated one, indicating a higher crystalline quality of the evaporated film. We remark that the quality factors in samples \#7 were obtained after optimizing the sputtering process -- for details of the samples, also their surface roughness and atomic force micrographs, see the Supplementary materials.\par

A short dip in HF post fabrication reduces the TLS loss, doubling the $Q_i$ achieved at low photon numbers, as shown in Fig.~\ref{fig:resonators_sputtered_andHF} and Table~\ref{tab:Sample-info-DCA}. Since it took us about \SI{2.5}{\hour} to place the HF-treated samples in vacuum (inside the dilution refrigerator's mixing chamber), we anticipate the regrowth of aluminum oxide at the MA interface; while it takes longer for the silicon oxide on the substrate to fully regrow~\cite{Verjauw2021}.  Therefore, we think that the improvement in the loss after the HF treatment is mainly due to the reduced contribution of the SA interface, while a minor improvement comes from the reduced loss at the MA interface: as discussed above, the partial etch of Al can help with removing the adsorbates on the aluminum surface. \par
A subsequent passivation of the silicon substrate by HMDS in sample \#7c slightly degrades $Q_i$ compared to \#7b, although $F\delta_\text{TLS}^0$ is still below one half that of the untreated sample \#7a, see Table~\ref{tab:Sample-info-DCA}. We note that due to the decreased resonance linewidth of the higher-quality resonators, single-photon levels are not reached in the measurement power span. Therefore, in Table~\ref{tab:Sample-info-DCA}, we compare the $Q_i$ of the resonators at $\langle n \rangle \sim 10$. The decreased TLS loss of the HF-treated samples becomes comparable to the other losses $\delta_{other}$, resulting in the low-power $Q_i$ being limited by the other loss mechanisms as well.
%

\section*{\uppercase{Conclusion}}
In conclusion, our investigations show that concerning the loss at the substrate-metal interface, the surface treatment with SC1 and HF is the most effective method to reduce the associated TLS loss in superconducting Al-based CPW resonators fabricated on Si substrates. ToF-SIMS characterization shows one order of magnitude higher intensity of oxygen at the SM interface if the native oxide is not removed from the substrate. In comparison to the standard sample, this leads to an increase in the TLS loss by 5 times. In addition, we found that avoiding preheating of the substrate prior to deposition can increase the TLS loss by about 60\%. \par 
The in-situ oxidation of the Al film led to a marginal increase in the loss; however, observation of voids in the Al film and residues in the gap suggests that this step is necessary for not facing random and unwanted open circuits in the device. Finally, despite the higher pre-deposition base pressure and temperature, the TLS loss of the sample with short pumping was still comparable to the standard process, reaffirming the insignificance of the bulk material quality and composition compared with the surfaces and interfaces as the source of TLS loss~\cite{Wenner2011}. \par  
With regard to the substrate-air -- and to a lower degree the metal-air -- interface, a post-fabrication HF dip of resonator chips reveals that removing the surface oxides of Al and Si can decrease the TLS loss by about 3 times. As the effect of the oxide strip is temporary, until an oxide has regrown, it is desirable to extend this effect for practical applications. We show that immediate passivation of the Si by HMDS using vapor deposition can slightly degrade the $Q_i$ achieved after the HF dip, although still keeping the loss well below that of an untreated sample. Further studies into the longevity of the protective effect of the HMDS layer are necessary. 

\section*{Acknowledgments}
This work was funded by the Knut and Alice Wallenberg (KAW) Foundation through the Wallenberg Centre for Quantum Technology (WACQT) and by the EU Flagship on Quantum Technology HORIZON-CL4-2022-QUANTUM-01-SGA project 101113946 OpenSuperQPlus100. The authors acknowledge the use of the Nanofabrication Laboratory at Chalmers University of Technology. Financial support from Swedish Research Council (VR) and Swedish Foundation for Strategic Research (SSF) for the access to ARTEMI, the Swedish National Infrastructure in Advanced Electron Microscopy (2021-00171 and RIF21-0026) is also acknowledged. The XPS, ToF-SIMS and TEM measurements were performed at the Departments of Industrial and Materials Science, Chemistry and Chemical Engineering, and the Chalmers Material Analysis Laboratory (CMAL), respectively. The authors acknowledge valuable feedback and comments from Sandoko Kosen, Henrik Frederiksen, Niclas Lindvall, Daniel Pérez Lozano, and the fabrication team at Quantum Technology laboratory (QT), Chalmers. 

\section*{Author Contributions}
Janka Biznárová and Lert Chayanun have contributed equally to the work.\par

\textbf{Conceptualization:} A.\,F.\,R. and Jonas\, B. \textbf{Fabrication and microwave measurements:} Janka\,B. and L.\,C. \textbf{Characterization and analysis:} Janka\,B., L.\,C., L.\,Z., A.\,N., P.\,M., M.\,R, E.\,T., A.\,F.\,R. \textbf{Funding:} Jonas\,B. \textbf{Supervision:} A.\,F.\,R., Jonas\, B. \textbf{Writing -- original draft:} L.\,C. Janka\, B., A.\,F.\,R., Jonas\,B. \textbf{Writing -- review \& editing:} all authors.



\newpage
\baselineskip13pt

\section*{{Supplementary Materials}}

\maketitle 
\section*{{Sputtering parameter optimization}}

In samples \#7a-12, we investigated the influence of varying deposition parameters of the thin aluminum film.   In contrast to the previous samples, these films were sputtered, and SC1 was omitted.
We kept the post-deposition fabrication procedure, starting from the in-situ oxidation, consistent with samples \#1-6.
We present the $Q_i$ values and the extracted loss parameters in Table~\ref{tab:SamplesDCAsup} and Fig.~\ref{fig:supp_DCA}.

\begin{itemize}
\item In samples \#8, \#7a, and \#9a, we varied the sputtering rate by changing the forward power applied to the DC magnetron used to ignite the Ar plasma. 
\item In sample \#10, we preheated the Si substrate to \SI{100}{\celsius} for \SI{10}{\min} prior to deposition, in contrast to the remaining samples whose deposition was started at room temperature (RT), with the intent to promote larger grain sizes. 
\item In samples \#11 and \#12, we heated the Si wafer in the vacuum of the deposition chamber at \SI{700}{\celsius} for \SI{25}{\min} instead of the standard \SI{300}{\celsius} preheat. We also cleaned the substrate of \#12 by SC1, in addition to the HF dip.  
\item We repeated the post-fabrication HF + HMDS treatment in sample \#9b, observing a 50\% decrease in $F\delta^0_{TLS}$ compared to sample \#9b, see Fig.~\ref{fig:supp_DCA_HF}.
\end{itemize}

This data indicates that there is an optimal intermediate deposition rate for decreasing TLS loss. Generally, increased deposition rates are desired to reduce the probability of trapping impurities in the film during growth. In addition, the higher kinetic energy of atoms sputtered at high power gives the atoms increased mobility. However, this can also enable implantation of the target material in the substrate. 

\setcounter{table}{0} 
\renewcommand{\thetable}{S\arabic{table}}
\begin{table}[b!]
\caption{\small{\label{tab:SamplesDCAsup} Summary of samples \#7a-12 where the Al film was sputtered. Sample \#7a is the closest in fabrication conditions to the evaporated standard \#1, the only differences being the sputtered deposition and the omission of SC1. The table includes the average $Q_\text{i}$ at $\langle n \rangle \sim 1$ , $\beta$, $F\delta_\text{TLS}^0$ and $\delta_\text{other}$ for the TLS model fit for every sample, as shown in Fig.~\ref{fig:supp_DCA}. }}

\small
\footnotesize
\renewcommand{\arraystretch}{1.1}
\centering\begin{tabular}{l c c c c c c }
\\[-1.8ex]
\hline \hline
\\[-1.8ex]
     & \textbf{Sample 8} & \textbf{Sample 7a} & \textbf{Sample 9a} & \textbf{Sample 10} & \textbf{Sample 11} & \textbf{Sample 12} \\
     & \textbf{(\#8)} & \textbf{(\#7a)} & \textbf{( \#9a)} & \textbf{(\#10)} & \textbf{(\#11)} & \textbf{(\#12)} \\[3pt]
   Variation & \SI{0.3}{\nano\meter\per\second} & \SI{1}{\nano\meter\per\second} & \SI{1.4}{\nano\meter\per\second} & \SI{100}{\celsius} dep. & \SI{700}{\celsius} anneal & \SI{700}{\celsius} anneal + SC1 \\[3pt]
   \hline \\[-1.8ex]
   
   $Q_\text{i}$\,($\langle n \rangle \sim 1$) & $0.40\times10^6$ & $1.0\times10^6$ & $0.58\times10^6$ & $0.32\times10^6$ & $0.48\times10^6$ & $0.38\times10^6$ \\[3pt]
			
   $\beta$ & $0.29$ & $0.36$ & $0.23$ & $0.26$ & $0.24$ & $0.21$ \\[3pt]

   $F\delta_\text{TLS}^0$ & $2.2\times10^{-6}$ & $0.86\times10^{-6}$ & $1.5\times10^{-6}$ & $2.9\times10^{-6}$ & $2.0\times10^{-6}$ & $2.6\times10^{-6}$ \\[3pt]

   $\delta_{other}$ & $1.6\times10^{-7}$ & $1.5\times10^{-7}$ & $1.3\times10^{-7}$ & $1.9\times10^{-7}$ & $2.0\times10^{-7}$ & $1.8\times10^{-7}$ \\[2pt]
\hline \hline
   
\end{tabular}

\end{table}

Sputtering Al at elevated substrate temperatures is common in CMOS industry \cite{Plummer_1999_book} to increase surface mobility and improve step coverage. In our case, however, deposition at \SI{100}{\celsius} resulted in a higher TLS loss (\#10). This may be caused by the significant increase in the film's roughness, see Fig.~\ref{fig:AFM}(c) and Table~\ref{tab:AFM}.

In samples \#11 and \#12, the substrate underwent annealing temperatures of above \SI{700}{\celsius} for surface reconstruction to achieve a better atomic transition between the metal and substrate \cite{Fritz2019,McSkimming2017}. However, this increased the TLS loss of these samples, which may have been caused by the deterioration of the residual hydrogen monolayer on the Si substrate after the HF treatment, and a subsequent re-oxidation of the Si surface. We also observe an increase in the Al surface roughness in the sample annealed to \SI{700}{\celsius}, Fig.~\ref{fig:AFM}(d), compared to the sample preheated to \SI{300}{\celsius}, Fig.~\ref{fig:AFM}(b). 

\setcounter{figure}{0}  
\renewcommand{\thefigure}{S\arabic{figure}}

\begin{figure}[t]
    \centering
    \includegraphics[width=8.5cm]{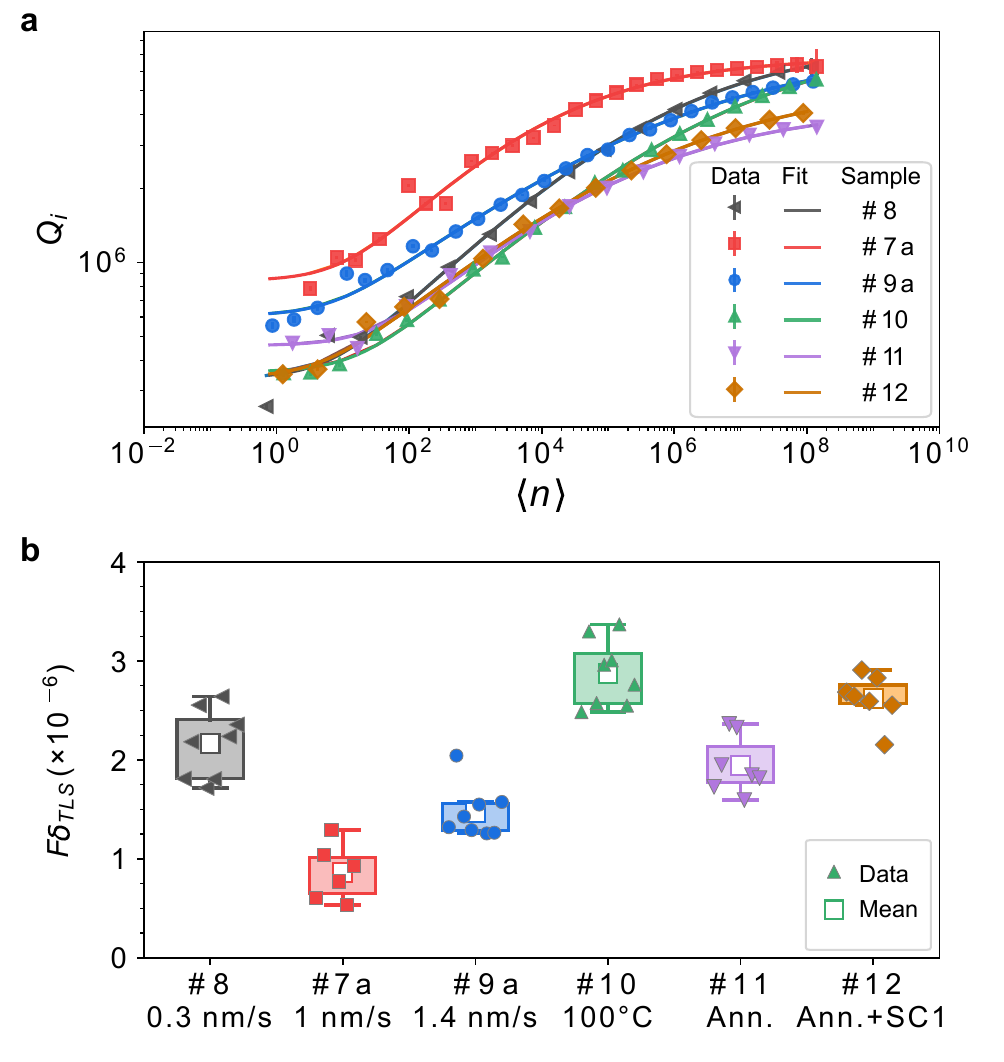}
    \caption{\small{\label{fig:supp_DCA} Internal quality factor ($Q_\text{i}$) and losses from the TLS model fitting for the samples with sputtered Al film listed in Table~\ref{tab:SamplesDCAsup}. \textbf{a)} $Q_\text{i}$ as a function of the circulating power. The resonance frequency is $\sim$~\SI{4.8}{\giga\hertz}. \textbf{b)} Box plot of TLS loss from the fitting with the TLS loss model for all of the resonators on each sample. Each data-point represents the loss of an individual resonator.
    }}
\end{figure}

\subsection*{Aluminum surface roughness}

\begin{figure}[t!]
    \centering
    \includegraphics[width=8.5cm]{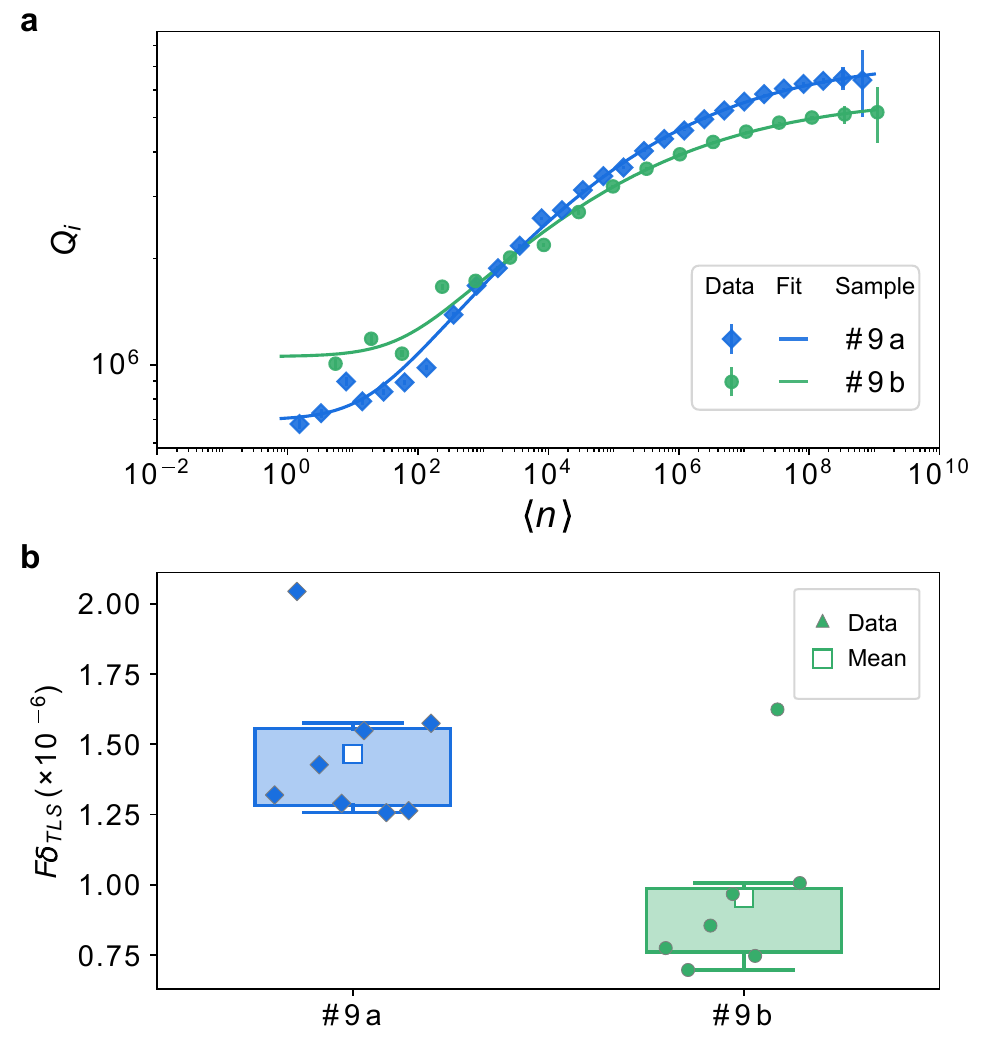}
    \caption{\small{\label{fig:supp_DCA_HF} Internal quality factor ($Q_\text{i}$) and losses from the TLS model fitting for two samples \#9a and \#9b with sputtered Al film. \textbf{a)} $Q_\text{i}$ as a function of the average number of photons for the resonators with  resonance frequency $\sim$~\SI{4.1}{\giga\hertz}. \textbf{b)} Box plot of TLS loss from  fitting to the TLS loss model for all of the resonators on each sample. Each data-point represents the loss of an individual resonator. Both chips originate from the same wafer. Sample \#9a is the control, with the process variations listed in Table~\ref{tab:SamplesDCAsup}. Sample \#9b has received treatment by both HF and HMDS post fabrication, achieving an average $F\delta^0_{TLS}$ of \SI{9.5e-7}{}, $\delta_{other}$ of \SI{2.2e-7}{}, $\beta$ of 0.18, and an average low-power $Q_i$ of \SI{9.1e5}{}.}}
\end{figure}

\begin{figure}[t]
    \centering
    \includegraphics[width=13cm]{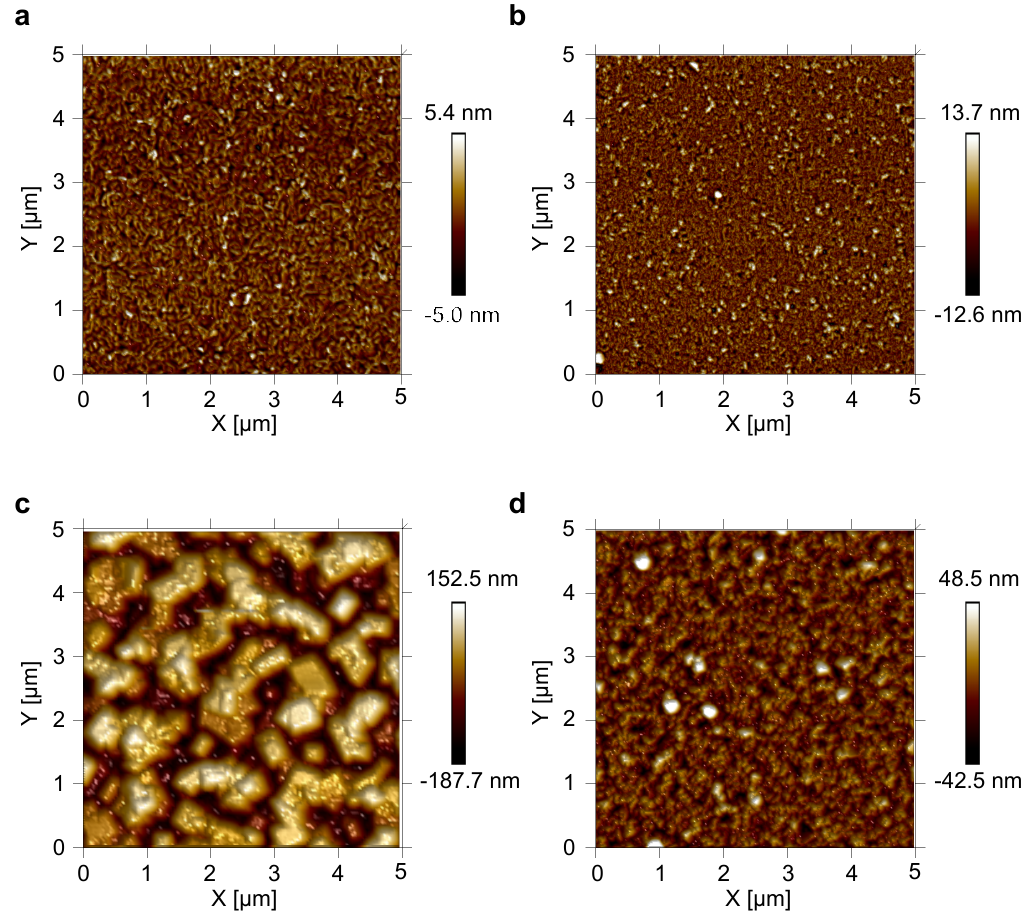}
    \caption{\small{\label{fig:AFM} AFM of the top surface of Al films deposited on Si using four different sets of parameters summarized in Table~\ref{tab:AFM}. 
    }}
\end{figure}

We used atomic force microscopy (AFM) to quantify the surface roughness of the Al films deposited on Si using different approaches. We obtained the quadratic mean roughness $R_q$, also known as the root mean square (RMS) roughness, by scanning a $5 \times \SI{5}{\micro \meter}$ area of the Al surface, shown in Fig.~\ref{fig:AFM}. 

\begin{table}[b!]
\caption{\small{\label{tab:AFM} Surface roughness values of different samples obtained from AFM shown in Figure~\ref{fig:AFM}, alongside the corresponding Al film deposition parameters.}}

\small
\footnotesize
\renewcommand{\arraystretch}{1.1}
\centering\begin{tabular}{c c c c }
\\[-1.8ex]
\hline \hline
\\[-1.8ex]
     Figure~\ref{fig:AFM} panel &  Substrate treatment  & Deposition method & $R_q$ \\[3pt]
     \hline \\[-1.8ex]
   \textbf{a} & \SI{300}{\celsius}, \SI{10}{\minute} & Evaporation, RT & \SI{1.4}{\nano\meter} \\[3pt]
   \textbf{b} & \SI{300}{\celsius}, \SI{10}{\minute} & Sputtering, RT & \SI{3.2}{\nano\meter} \\[3pt]
	\textbf{c} & \SI{300}{\celsius}, \SI{10}{\minute} & Sputtering, \SI{100}{\celsius} & \SI{58.9}{\nano\meter} \\ [3pt]
    \textbf{d} & \SI{700}{\celsius}, \SI{25}{\minute} & Sputtering, RT & \SI{11.3}{\nano\meter} \\[3pt]		
\hline \hline
   
\end{tabular}
\end{table}
The samples imaged here have experienced the full resonator fabrication process. The fabrication steps were consistent with the standard process outlined in the Method section, with the exception of the omission of SC1 in these samples, and the varying substrate heating and film deposition parameters. The deviating fabrication conditions of these samples are listed in Table~\ref{tab:AFM} alongside the obtained $R_q$ values. 

Samples \textbf{a} and \textbf{b} have undergone identical substrate treatments, followed by Al deposition at an equal deposition rate -- however, in sample \textbf{a}, the Al was deposited by evaporation, and in \textbf{b} by sputtering. The $R_q$ of the sputtered film is approximately twice that of the evaporated film.
As the TLS loss of resonators fabricated on these two films is comparable, we conclude that at the loss rates that our current devices are experiencing, the potential increase of the MA loss induced by the doubling of the MA surface roughness is negligible.

Sample \textbf{c} was deposited on a substrate heated to \SI{100}{\celsius} prior to sputtering. Here the increase of film roughness is significant, to the level of being noticeable by the naked eye -- under ambient light, the finish of this film appears white, as opposed to the usual mirror-like finish of an Al thin film. The size of the grains in Fig.~\ref{fig:AFM}(c) is noticeably larger than in the other samples. 

The surface roughness in sample \textbf{d}, deposited using the same conditions as sample \textbf{b} but on a substrate heated to \SI{700}{\celsius} prior to an overnight pump-down and subsequent deposition, increased 3.5 times compared to \textbf{b}. 

\subsection*{Defects in the Al film oxidized in ambient air}

\begin{figure}[b!]
    \centering
    \includegraphics[width=7.5cm]{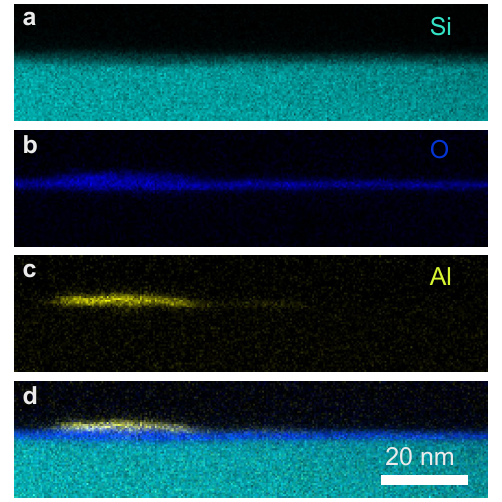}
    \caption{\small{\label{fig:supp_EDS_patch} EDS of a flake in an area of sample \#5 where the Al has been etched. \textbf{a)}, \textbf{b)}, \textbf{c)} and \textbf{d)} show the maps of Si, O, Al, and a composite image of the three elements, respectively.}}    
\end{figure}

We use STEM EDS to study the composition of the flakes left behind on the Si surface after etching the Al on sample \#5, which was not oxidized in-situ. We show the presence of Al and Si, as well as a locally thicker oxide layer in the area of the flake, in Fig.~\ref{fig:supp_EDS_patch}.

\begin{figure}[H]
    \centering
    \includegraphics[width=15cm]{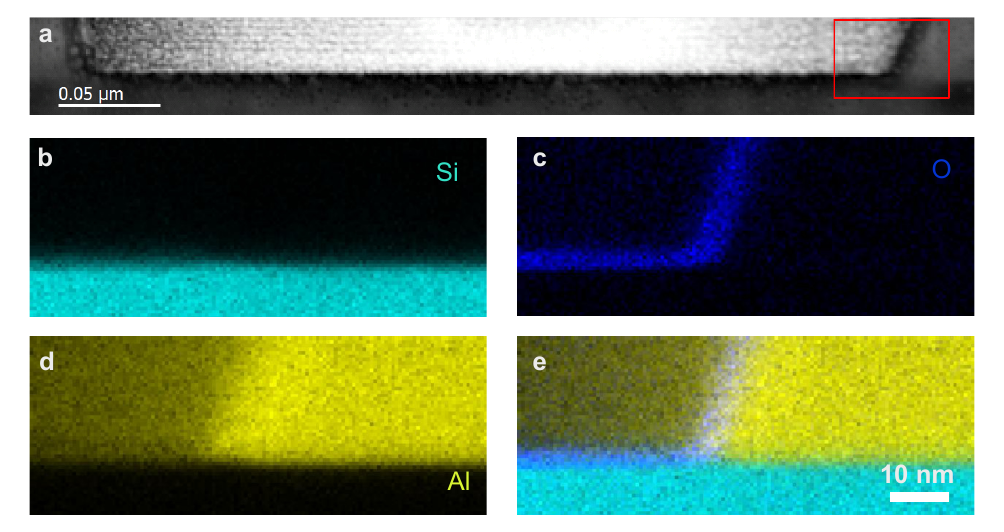}
    \caption{\small{\label{fig:supp_EDS_void} STEM \textbf{(a)} and EDS \textbf{(b--e)} of a void in the Al film of sample \#5. \textbf{b)}, \textbf{c)}, \textbf{d)} and \textbf{e)} show the detected Si, O, Al, and a composite image of the three elements, respectively.}
    }
\end{figure}

We also study the elemental distribution around the voids formed in the Al film in Fig.~\ref{fig:supp_EDS_void} using EDS. We detect a thicker oxide on the Al surface one the walls of the void ($\sim\SI{6.5}{\nano\meter}$) than in other areas of the film where voids are not in immediate vicinity ($\sim\SI{5}{\nano\meter}$). This supports the hypothesis that the inconsistent oxidation in ambient air as opposed to the in-situ oxidation in pure oxygen gives rise to the formation of these defects.

\bibliography{refs.bib}
\bibliographystyle{unsrt}

\end{document}